\documentstyle[11pt,moriond,epsfig]{article}

\def\Journal#1#2#3#4{{#1} {\bf #2}, #3 (#4)}

\def\PLB{{\em Phys. Lett.}  B}

\def\PRD{{\em Phys. Rev.} D}

\def\be{\begin{equation}}
\def\ee{\end{equation}}
\def\bea{\begin{eqnarray}}
\def\eea{\end{eqnarray}}

\begin{document}
\vspace*{4cm}
\title{PROMPT PHOTONS\\ AND\\ DEEPLY VIRTUAL COMPTON SCATTERING\\ AT HERA}

\author{ I. GRABOWSKA-BO\L D  }

\address{University of Mining and Metallurgy, Faculty of Physics and Nuclear Techniques, \\
Al. Mickiewicza 30, Cracow, Poland\\
E-mail: iwona.grabowska@desy.de\\
(on behalf of the ZEUS and H1 Collaborations)}

\maketitle\abstracts{
Two results in $ep$ physics using HERA data are discussed: a measurement of the effective transverse parton momentum in the proton by studying the production of prompt-photons in photoproduction events in the ZEUS collaboration and differential cross sections for Deeply Virtual Compton Scattering obtained by both the H1 and ZEUS collaborations.  
}

\section{Introduction}
\begin{figure}
\begin{minipage}{8.5cm}
\begin{tabular}{cc}
\centering
\epsfig{file=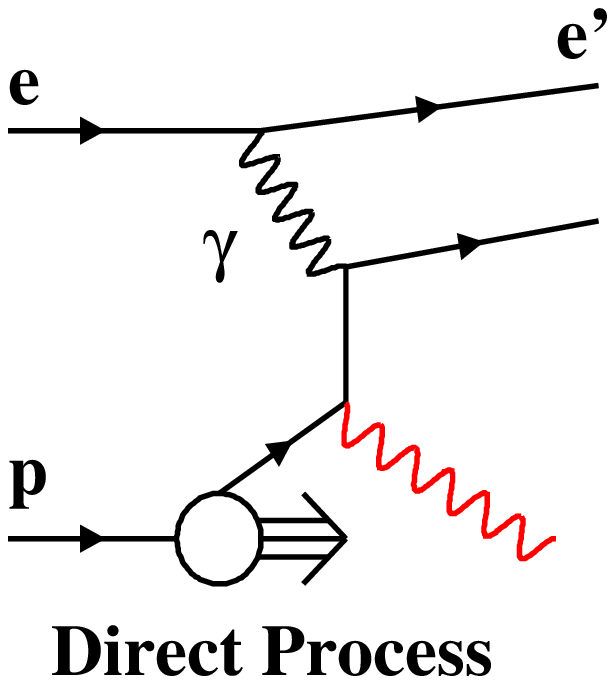,height=4.1cm} &
\epsfig{file=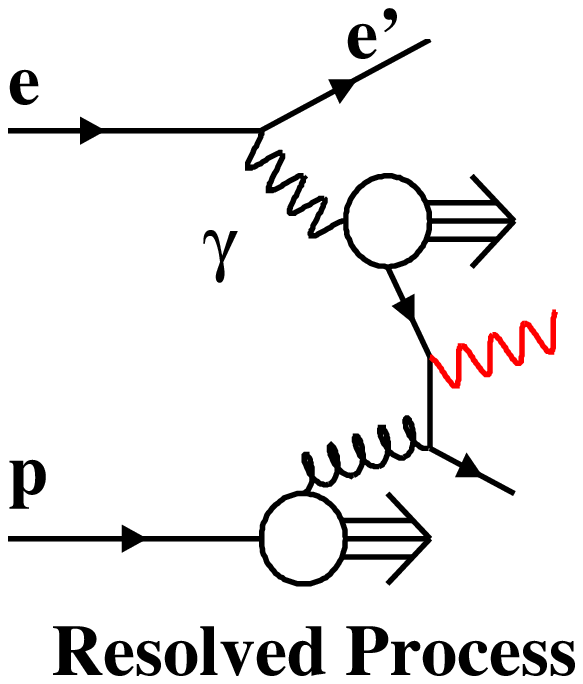,height=4.1cm} 
\end{tabular}
\caption{Diagrams for the prompt-photon process in the direct and resolved photoproduction.\label{fig:prompt1}}
\end{minipage}%
\begin{minipage}{1cm}
\hspace*{1cm}
\end{minipage}%
\begin{minipage}{6cm}
\centering
\epsfig{file=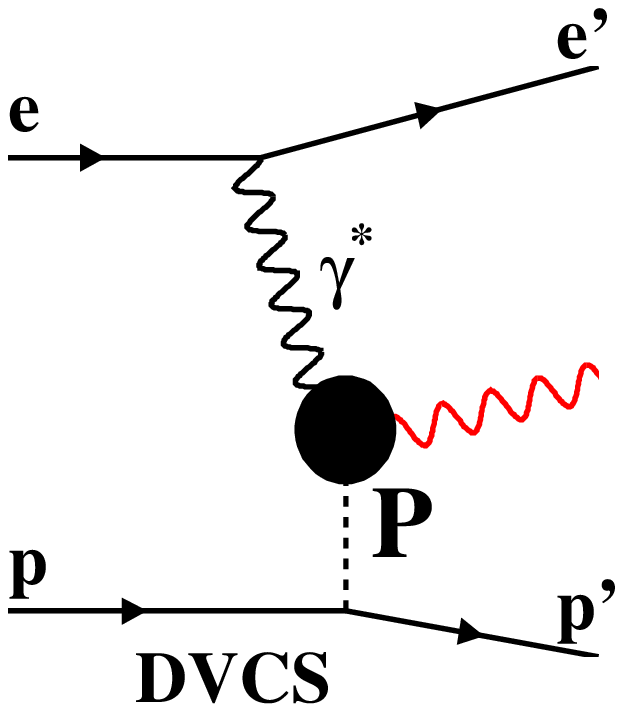,height=4.1cm} 
\caption{A diagram for the DVCS process.\label{fig:dvcs1}}
\end{minipage}
\end{figure}

 Two types of real-photon production processes in $ep$ collisions are studied at HERA: prompt-photon production in photoproduction and Deeply Virtual Compton Scattering (DVCS) in deep inelastic scattering (DIS). These reactions in high energy collisions are of interest because they allow the investigation of parton level interactions while minimizing the complication of the hadronisation effect.\\
\indent Prompt-photon production in photoproduction is the production of a real photon directly from the hard interaction of a quasi-real photon with a proton. Two kinds of processes can be defined at leading order of QCD: {\em direct} (where the entire quasi-real photon takes part in the interaction) and {\em resolved} (where the incoming photon fluctuates into a hadronic system prior to the interaction) as depicted in Fig.\ref{fig:prompt1}. ZEUS has used the direct prompt-photon process to assign an effective transverse parton momentum in the proton, $\langle k_T \rangle$.\\
\indent The second process discussed is DVCS, in which a real photon is produced in the diffractive scattering of a virtual photon off a proton, as shown in Fig.\ref{fig:dvcs1}. The diagram is similar to vector meson production but free of the complications introduced by the vector mesons wave functions. The simplicity of the process may give access to the skewed parton distributions which describe two-parton correlations in the proton. Both the H1 and ZEUS collaborations have already seen first signals for DVCS and now report on the differential cross section measurements.\\

\section{Prompt Photons in Photoproduction}\label{sec:prompt}
The ZEUS analysis \cite{pp} of the prompt-photon process is based on data taken at HERA in the 1996-97 running period with energies of 27.5 GeV and 820 GeV for positrons and protons, respectively.\\
\indent The determination of $\langle k_T \rangle$ is facilitated by the use of the direct process, thereby avoiding any additional contribution to the effective transverse parton momentum in a proton from the resolved photon. At leading order in photoproduction, the Compton process is the only direct prompt-photon process.\\
\indent To determine $\langle k_T \rangle$ the data were compared to different predictions from PYTHIA \cite{pythia}. Within PYTHIA it is possible to vary the intrinsic smearing on the transverse momentum of the partons in an incoming hadron. This smearing is imposed in addition to the effect of parton showering. The effective transverse parton momentum is the combination of intrinsic and parton shower components.\\ 
\indent To find the intrinsic momentum the distribution of the momentum component of the photon perpendicular to the jet direction was compared to the predictions of PYTHIA for different values of the parameter $k_0$, which is the variable directly connected to the mean absolute value of the intrinsic parton momentum in the proton $\langle k_T^{intr} \rangle$. For different $k_0$ values the fit to the data was performed to determine the optimal value of $\langle k_T^{intr} \rangle$, which was found to be \mbox{$1.25 \pm 0.41 ^{+0.15}_{-0.28}$ $\rm{GeV}$}. The parton-shower contribution to $\langle k_T \rangle$ was found to be approximately $1.4$ $\rm{GeV}$. The total $\langle k_T \rangle$ was found to be
\begin{center}
$\langle k_T \rangle = 1.69 \pm 0.18(stat.)^{+0.18}_{-0.20}(syst.)$ $\rm{GeV}$.
\end{center} 

\begin{figure}[h]
\begin{center}
\epsfig{file=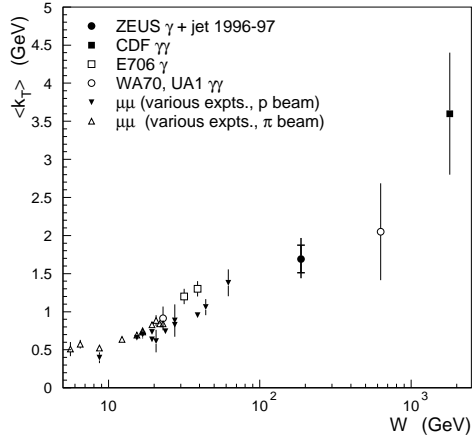,height=6cm,bb=0 50 490 520, clip} 
\caption{The ZEUS measurement of $\langle k_T \rangle$, the effective mean transverse momentum of the partons in the proton,  compared with the results from other experiments as a function of the center-of-mass energy of the colliding particles.\label{fig:prompt4}}
\end{center}
\end{figure}

\indent Fig. \ref{fig:prompt4} shows the ZEUS result in comparison with the mean $\langle k_T \rangle $ values of other experiments, plotted as a function of the center-of-mass energy, $W$, of the incoming particles. 
The ZEUS result bridges a gap between low and high-energy measurements. Although different experimental methods have been employed, a clear trend for $\langle k_T \rangle$ to rise with increasing $W$ is evident. The ZEUS result is fully consistent with this trend.

\section{Deeply Virtual Compton Scattering}
\begin{figure}[h]
\centering
\begin{tabular}{cc}
\epsfig{file=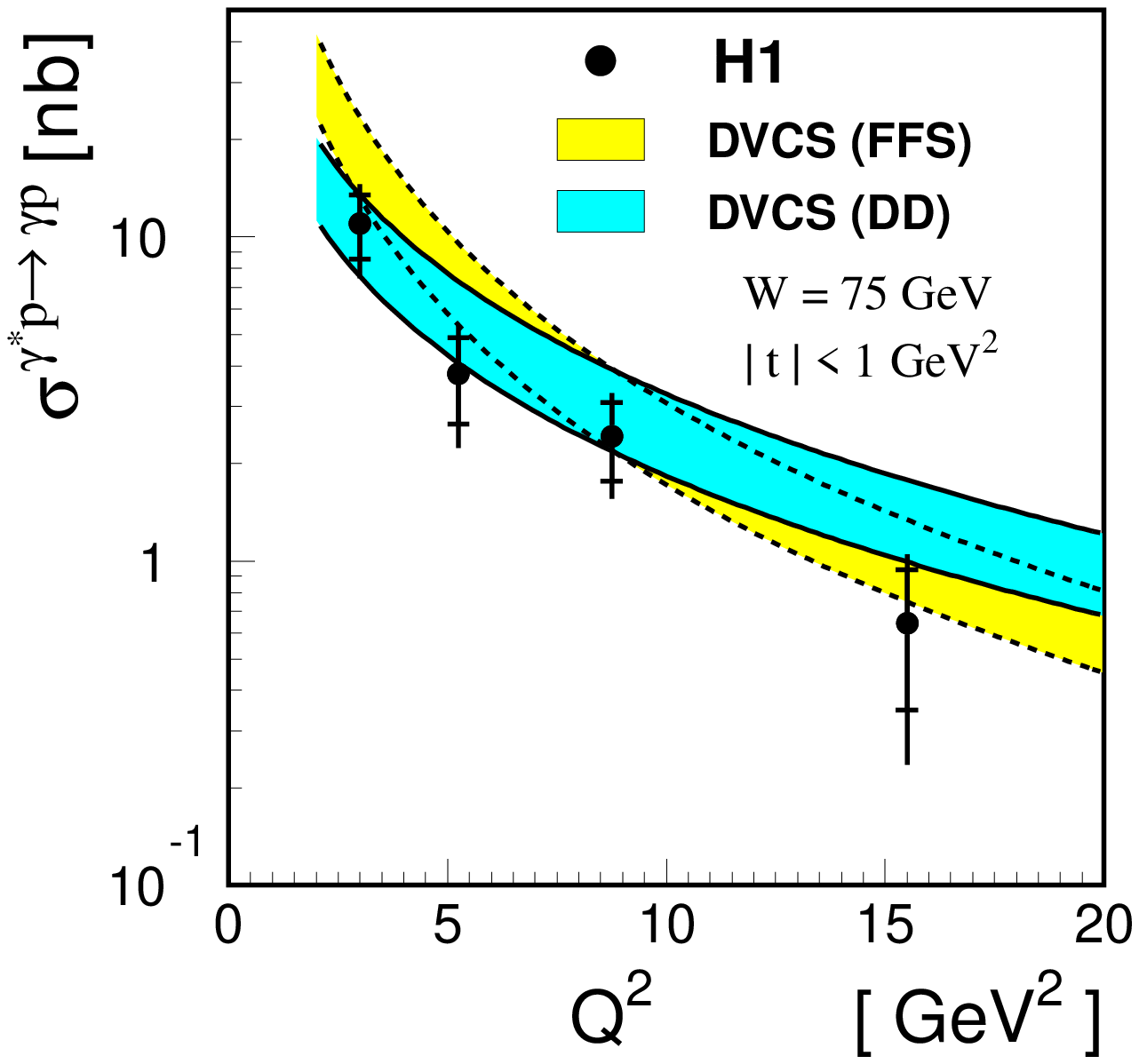,height=6cm,bb=110 240 480 585, clip} &
\epsfig{file=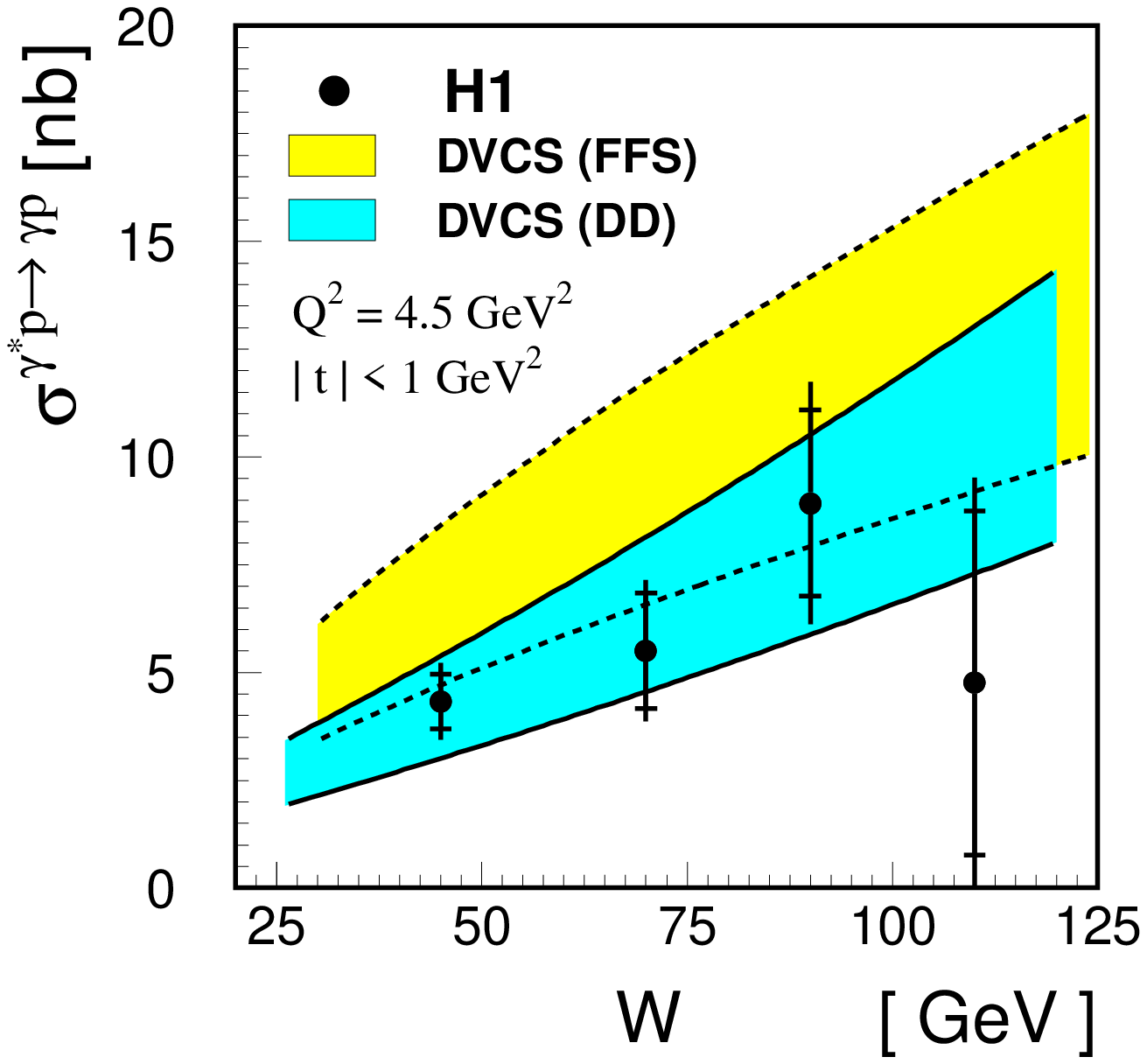,height=6cm,bb=110 240 480 585, clip} \\
\end{tabular}
\caption{The DVCS $\gamma ^*p$ cross sections as a function of $Q^2$ (left) and $W$ (right) measured by H1.\label{fig:dvcs_h1}}
\end{figure}

QCD-based models for DVCS were developed by Frankfurt, Freund and Strikman (FFS) \cite{ffs} and Donnachie and Dosch (DD) \cite{dd}. Both contain ``soft'' (nonperturbative) and ``hard'' contributions. The soft part in the FFS prediction is based on the aligned jet model, whereas reggeon and soft pomeron exchanges are considered in the DD prediction. The hard contribution in FFS is calculated in perturbative QCD and in DD is based on the dipole model. Since the theory doesn't predict the $t$-dependence for DVCS, where $t$ is the four momentum transfer squared at the proton vertex, it is assumed to be exponential, $e^{-b|t|}$, with $b$ laying in the range [4-10] $\rm{GeV^{-2}}$ for HERA data. Both H1 and ZEUS have developed Monte Carlo generators based on these calculations.\\
\indent The main source of the background in the DVCS analysis is the electromagnetic Bethe-Heitler (BH) process. Studies by ZEUS have shown \cite{dvcs_pat1} that it is possible to separate the DVCS signal from the BH background selecting events with a scattered lepton in the backward region of the detector and a photon in the central or forward part of the detector, where the direction is defined with respect to the incoming proton beam.\\
\indent Both the H1 and ZEUS collaborations used the same method to extract the DVCS signal. The selection requires two electromagnetic clusters in the detector with at most one track associated to one of the clusters. Details of both analyses are described elsewhere \cite{dvcs_h1,dvcs_pat2}. \\
\indent    To calculate the cross sections, the data have been corrected for acceptance, detector effects and initial state radiation. The range of the cross section measurement is slightly different in the two experiments.\\
\begin{figure}[h!]
\centering
\begin{tabular}{cc}
\epsfig{file=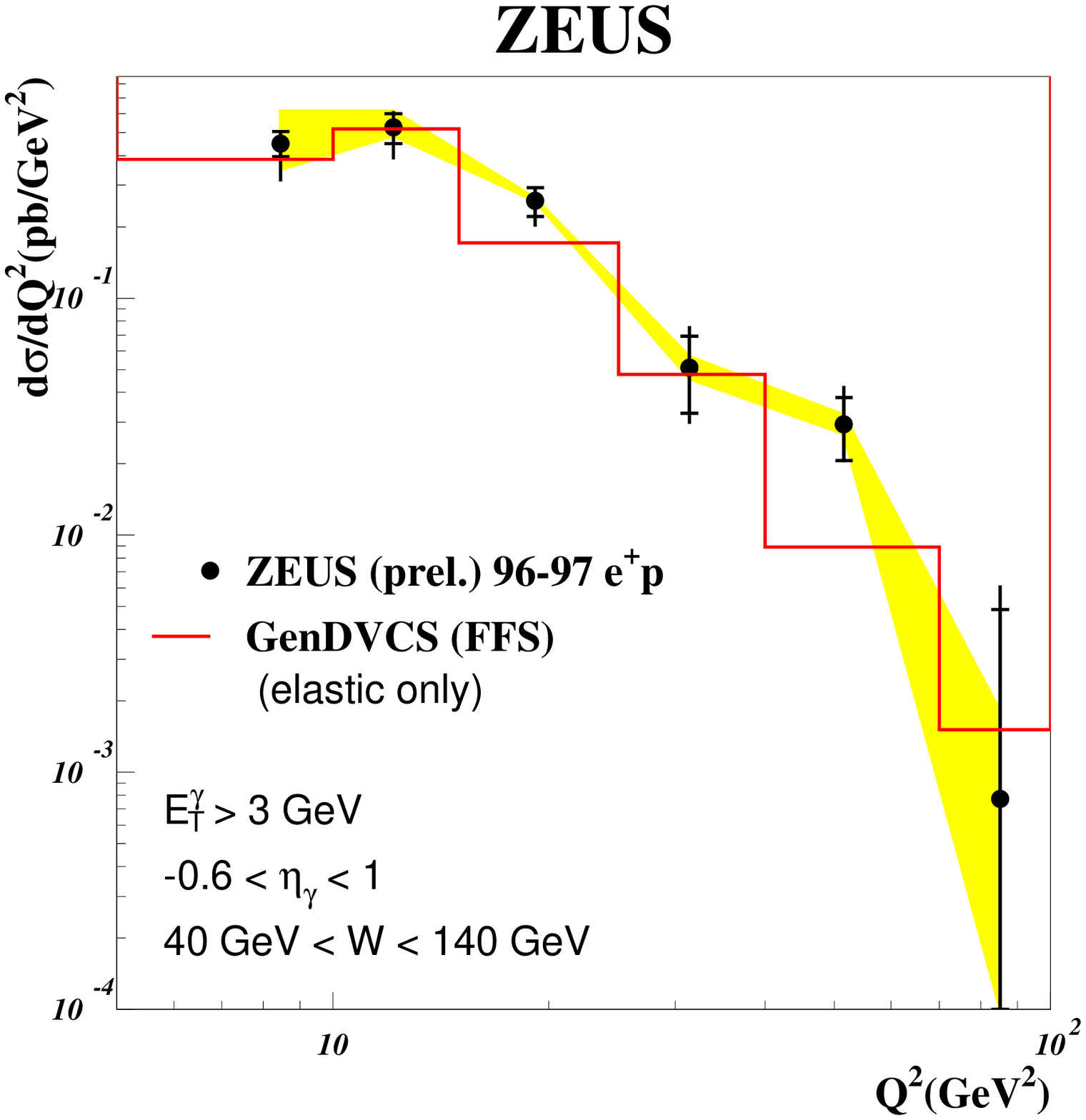,height=6cm} &
\epsfig{file=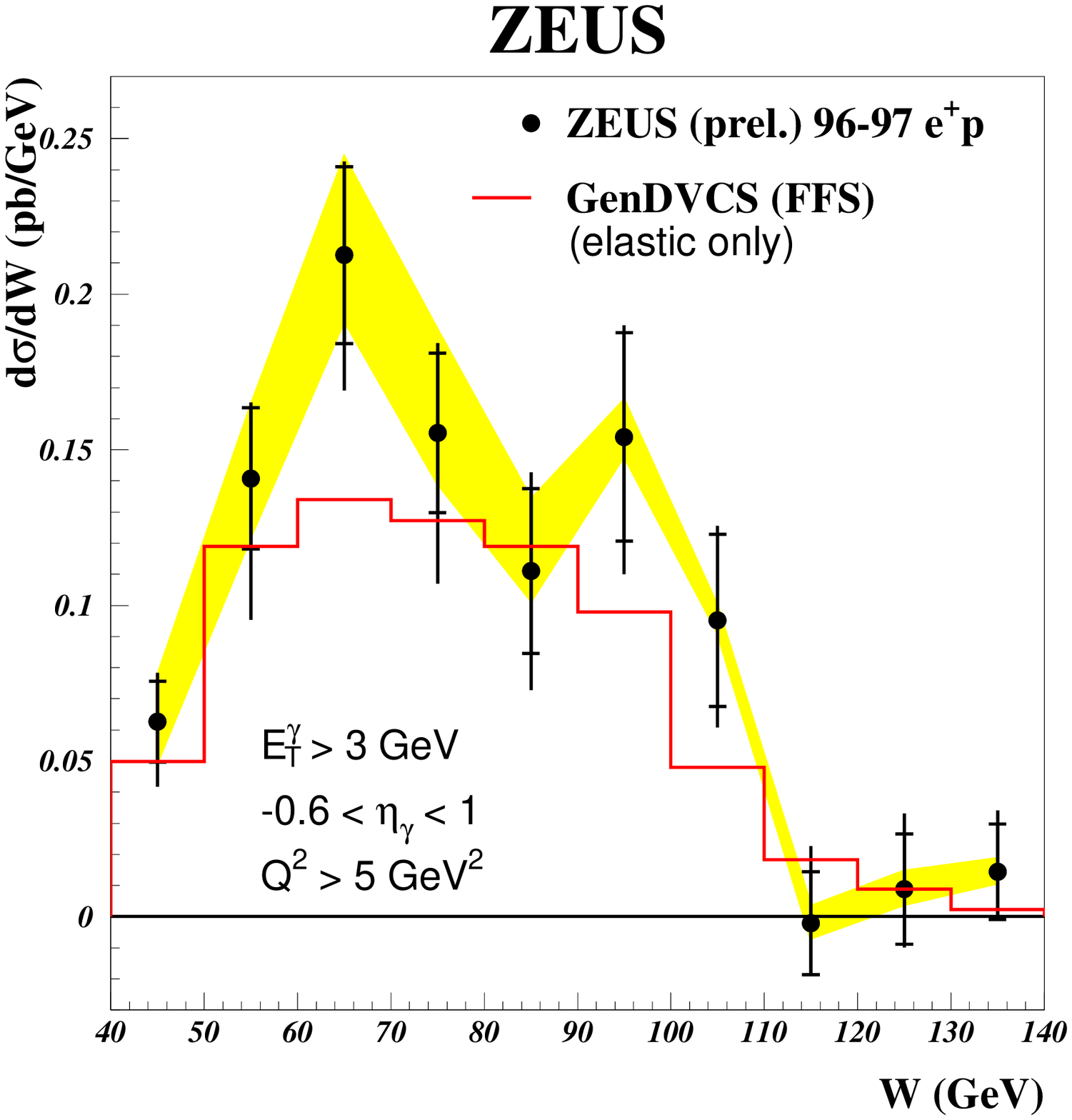,height=6cm} \\
\end{tabular}
\caption{The DVCS differential cross sections as a function of $Q^2$ (left) and $W$ (right) measured by ZEUS.\label{fig:dvcs_zeus}}
\end{figure}
\indent Fig.\ref{fig:dvcs_zeus} shows the differential cross sections as a function of $Q^2$ and $W$ obtained by ZEUS. The data are compared with the FFS prediction. The shaded band corresponds to the energy scale uncertainty which is correlated between bins and is not included to the systematic error. Good agreement between data and the QCD-based model is found. 20\% contribution from the proton-dissociation reaction  has not been subtracted from the data and affects the data normalisation. \\  
\indent In Fig.\ref{fig:dvcs_h1} the H1 $\gamma ^* p \rightarrow \gamma p$ cross sections are presented and compared to both the FFS and DD expectations.
The shaded bands represent the change in overall normalisation of the theoretical prediction coming from a variation of the $t$-slope in the range $b = 5-9$ $\rm{GeV^{-2}}$. Within the errors, the data are in agreement with both theoretical models.

\section{Summary}
The ZEUS collaboration has used the prompt-photon process in photoproduction to determine the effective transverse parton momentum in the proton, $\langle k_T \rangle$. This result is consistent with the rise with $W$ seen in other experiments.\\
\indent The H1 and ZEUS collaborations have seen the first signals for DVCS and now report on the cross section measurements. The results are consistent with the FFS and DD theoretical predictions. The $t$-dependence is still under investigation. The higher luminosity expected after the HERA upgrade will be a great advantage in the study of the DVCS process.
 
\section*{Acknowledgments}
I would like to thank the Organizers of the Moriond Conference for the financial support.
I also wish to express my gratitude to the DESY Directorate for the partial support.

\end{document}